\documentclass{superfri}
\usepackage{graphicx}
\usepackage[hidelinks]{hyperref}
\usepackage{amsmath}
\usepackage{multirow}
\usepackage{rotating}
\usepackage{makecell}

\begin{document}

\author{Alexandra~A.~Kirsanova\footnote{\label{susu}South Ural State University, Chelyabinsk, Russia}\orcidID{0000-0002-8447-6596}\and Gleb~I.~Radchenko\footnoteref{susu}\orcidID{0000-0002-7145-5630}\and Andrei~N.~Tchernykh\footnoteref{susu}$^{,}$\footnote{\label{cicese}CICESE Research Center Ensenada, Mexico}$^{,}$\footnote{\label{ISP RAS}Ivannikov Institute for System Programming of the RAS, Russia}\orcidID{0000-0001-5029-5212}
}

\title{FOG COMPUTING STATE OF THE ART: CONCEPT AND CLASSIFICATION OF PLATFORMS TO SUPPORT DISTRIBUTED COMPUTING SYSTEMS}

\maketitle{}

\begin{abstract}%
As the Internet of Things (IoT) becomes a part of our daily life, there is a rapid growth in the connected devices. A well-established approach based on cloud computing technologies cannot provide the necessary quality of service in such an environment, particularly in terms of reducing data latency. Today, fog computing technology is seen as a novel approach for processing large amounts of critical and time-sensitive data. This article reviews cloud computing technology and analyzes the prerequisites for the evolution of this approach and the emergence of the concept of fog computing. As part of an overview of the critical features of fog computing, we analyze the frequent confusion of the concepts of fog and edge computing. We provide an overview of fog computing technologies: virtualization, containerization, orchestration, scalability, parallel computing environments, as well as a systematic analysis of the most popular platforms that support fog computing. As a result of the analysis, we offer two approaches to classification of the fog computing platforms: by the principle of openness/closure of components and a three-level classification based on the provided platform functionality (Deploy-, Platform- and Ecosystem as a Service).

\end{abstract}

% -----------------------------------------------------------------------
\section*{Introduction}
\label{sec:intro}
Data is a major commodity today. Having more data and the ability to intelligently analyze it effectively creates significant value for data-managed enterprises \cite{Hagiu2020}. According to the International Data Corporation (IDC), the amount of digital data generated in 2010 exceeded 1 zettabyte \cite{Reinsel2011}. 2.5 exabytes of new data have been generated daily since 2012 \cite{McAfee2012}. Cisco estimates that there will be about 50 billion connected devices by 2020 \cite{D.2011}. These connected devices form the Internet of Things (IoT) and generate a vast amount of data in real-time. Modern mobile networks are already being designed considering the loads that arise in the transmission and processing of such astronomical volumes of data.

Within the cloud computing concept, most of the data that requires storage, analysis, and decision making is sent to data centers in the cloud \cite{Ravandi2017}. As the data volume increases, moving information between an IoT device and the cloud may be inefficient or even impossible in some cases due to bandwidth limitations or latency requirements. As time-sensitive applications (such as patient monitoring, autopilot vehicles, etc.) become more common, the remote cloud will not be able to meet the need for ultra-reliable communications with minimal delay \cite{Zhang2020}. Moreover, some applications may not be able to send data to the cloud because of privacy issues.

To solve the challenges of applications that require high network bandwidth, access to geographically distributed data sources, ultra-low latency, and localized data processing, there is a specific need for a computing paradigm that provides a one-size-fits-all approach to the organization of computing, both in the cloud and on computing nodes closer to connected devices. The concept of Fog computing has been proposed by industry and academia to bridge the gap between the cloud and IoT devices by providing computing capabilities, storage, networking, and data management at network nodes located close to IoT devices \cite{Bonomi2012, OpenFogConsortiumArchitectureWorkingGroup2017}. The research community has proposed several computing paradigms to address these problems, such as edge computing, fog computing, and dew computing. A common feature of these concepts is the use of distributed heterogeneous systems that provide highly scalable clusters of computing nodes located near (either networked or geographically) to data sources. In this review, we provide an analysis of the most popular platforms that support fog computing solutions. Based on this analysis, we propose two approaches to classify fog computing platforms: by the principle of openness/closure of components and as a three-tier classification based on the provided platform functionality (Deploy-, Platform- and Ecosystem as a Service). 

The article is organized as follows. Section 1 discusses cloud computing as the basis for new computing concepts, prerequisites for the emergence, and key characteristics of cloud computing. Section 2 is devoted to fog and edge computing, their origins, definition, and critical characteristics. Section 3 discusses technologies that support fog computing, including virtualization, orchestration, security, and computation scalability issues. Section 4 provides an overview of fog computing platforms: private, public, open-source, and proposes a classification of fog platforms. In Section 5 we focus on the current challenges faced by the fog computing researchers.  In conclusion, we summarize the results obtained in the context of this study and indicate directions for further research. 
 
\section{Cloud computing as a basis for new computational concepts} \label{Cloud computing as a basis for new computational concepts}
\subsection{The prerequisites for cloud computing} \label{The prerequisites for cloud computing}
The utility computing concept, originating in the 1960s, is considered the earliest ancestor of cloud technologies \cite{Feeney1974, Webb2000}. This concept was not generally adopted until the 90s due to the technical constraints of the deployment and use of this architecture \cite{Armbrust2010, Brynjolfsson2010, Hannabuss2009, Liu2005, Madsen2013, Webb2000}. Improvements in network technology and data transfer rates in the mid-'90s led to a new round of research in utility computing in the framework of the grid computing concept \cite{Foster2011, Hannabuss2009, Liu2005}. These shortcomings have led to further evolutionary development and the emergence of cloud computing, which often uses the grid computing model to expand computing resources \cite{Kumar2015}. 

\subsection{Key Features of Cloud Computing} \label{Key Features of Cloud Computing}
Today, cloud computing systems have become widely used for Big Data processing, providing access to a wide variety of computing resources and a greater distribution between multi-clouds \cite{Radchenko2019}. This trend has been strengthened by the rapid development of the Internet of Things (IoT) concept. Virtualization via virtual machines and containers is a traditional way of organization of cloud computing infrastructure. Containerization technology provides a lightweight virtual runtime environment. In addition to the advantages of traditional virtual machines in terms of size and flexibility, containers are particularly important for integration tasks for PaaS solutions, such as application packaging and service orchestration.

The National Institute of Standards and Technology (NIST) published a definition of cloud computing, its main characteristics, and its deployment and maintenance models in 2011. Cloud computing has been defined as a model for enabling ubiquitous, convenient, on-demand network access to a shared pool of configurable computing resources (e.g., networks, servers, storage, applications, and services) that can be rapidly provisioned and released with minimal management effort or service provider interaction.

The NIST model comprises five essential characteristics, three service models, and four deployment models for clouds \cite{Mell2012}. The following key cloud deployment models can be identified: private cloud, public cloud, and hybrid cloud \cite{Eugene2013, Weinhardt2009}. 

A private cloud is deployed within a single organization, is available only to internal users, and does not share its resources outside the organization. The public cloud is developed by third parties and provides the resources to external users under the terms of the contract on the right of use. A hybrid cloud combines two types of deployment described above, which allows building a balance between private and public computing \cite{Eugene2013}.

Private clouds are commonly deployed as close to the end-user of the cloud as possible. That reduces the response time of the computing platform and increases the speed of data transfer between nodes of the system. However, a private cloud is tightly interconnected with the computing needs of its owner. Not every organization has enough resources to maintain its private cloud, which must meet the requirements for availability, reliability, and the law's requirements in the country where the cloud is located \cite{Hofmann2010, Sehgal2018}.

On the other hand, public cloud users often lack direct control over the underlying computing infrastructure. This can lead to several problems, including uncontrolled access by third parties to the private data hosted in a public cloud; blocking user servers that can be deployed on the same subnet with hosts banned in a particular country; the uncertainty of the quality of cloud resources as they are deployed on servers shared with third parties \cite{Hofmann2010}. It is also challenging to ensure a change of cloud provider, as it is necessary to solve the problem of migration and conversion of data and computing services.

These features of each type of deployment are the reason why cloud providers that provide clouds to private organizations often support the ability to create hybrid clouds \cite{Sotomayor2009}, which can be configured to a particular mode of operation, depending on the customer's requirements. This approach addresses data latency, security, and migration issues while maintaining the flexibility to customize computing resources for each task.

\subsection{Preconditions for new computing concepts} \label{Preconditions for new computing concepts}
Despite all the significant advantages guaranteed by public cloud platforms, problems that such approaches cannot effectively solve have emerged in the last five years. Thus, a large number of users of "smart" systems such as "smart home", "smart enterprise", "smart city" and other IoT solutions cannot always be satisfied with the quality of services provided by cloud solutions, in particular, due to the increase in the amount of data sent between the user/device and the cloud \cite{Hong2017}.

The emergence of the “smart” systems approach, populated with a variety of Internet-connected sensors and actuators, led to a revision of the architectural concept of data collection and analysis systems. The Internet of Things concept requires new approaches to storage solutions, fast data processing, and the ability to respond quickly to changes in the state of end devices \cite{Pinchuk2018, Proferansov2017, Yousefpour2019}. Also, the spread of mobile devices as the main platforms for client applications makes it difficult to transfer and process large amounts of data without causing problems with response delays due to the constant movement of mobile devices. 

As the amount of data sent between IoT devices, clients, and the cloud increases, problems associated with increased response time due to physical bandwidth limitations appear \cite{Madsen2013}. On the other hand, there are response time-sensitive applications and devices such as life support systems, autopilots, drones and others. Under these conditions, a remote centralized cloud has become unable to meet the ultra-low latency requirements \cite{Yousefpour2019}. Also, data transmission through multiple gateways and subnets raises the issue of sensitive data transmission \cite{Iorga2018}. 

In response to these problems, private enterprises and the academic community have raised the need to develop a computing paradigm that meets new concepts such as IoT \cite{Bonomi2012,Mahmood2018, Proferansov2017}. This paradigm had to fill the gap between the cloud and the end devices, providing computing, storage, and data transfer in intermediate network nodes closest to the end devices. Several paradigms have been developed and applied to solve this problem, including fog and edge computing \cite{Dar2019}. Each of these paradigms has its specific features, but all of them derive from a common principle - reducing time delays in data processing and transmission by moving computing tasks closer to the final device.

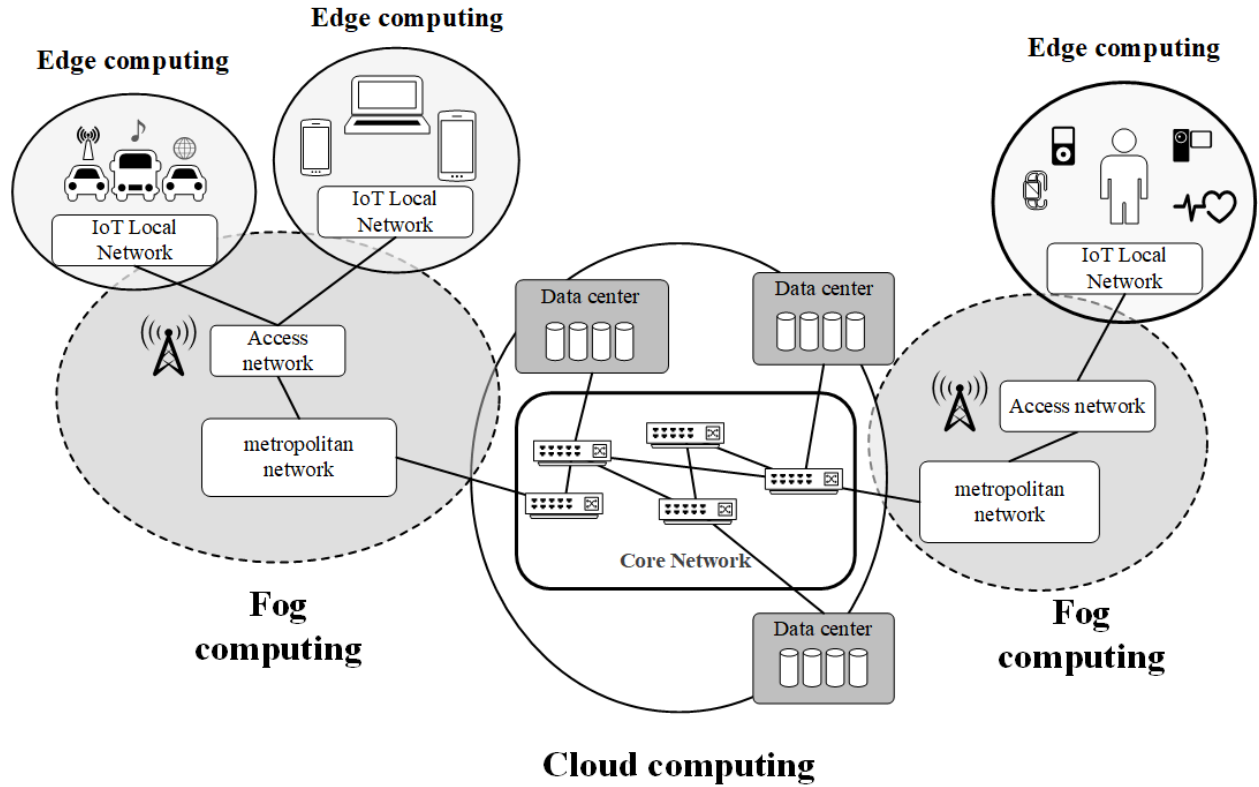
\fig{width=1\textwidth}{pic/Fig-Architecture}{Comparison of the infrastructure of fog computing and its related computing paradigms from the networking perspective \cite{Webb2000}}

\figref{pic/Fig-Architecture} shows a diagram of the relative distribution of computational resources defined by edge, fog and cloud computing concepts. Cloud computing is a separate data center (DC) or a network of data centers located far from the user but providing high computing capabilities. On the other hand, edge computing is located right at the edge of the computing system and provides small computing capabilities, but near the consumer of those resources. Fog computing is located between the edge of the network and the cloud data center, providing significant computing resources close to the end-user, which, on the other hand, is not comparable to the total amount of cloud computing resources but can be customized and scale depending on the objectives of the end-user. This article will consider Fog computing as a more general concept that includes the edge computing paradigm \cite{Hong2019}. 

\section{Fog and edge computing} \label{Fog and edge computing}
\subsection{History and definition} \label{History and definition}
In 1961 (see ref{tab-timeline}), John McCarthy spoke at the MIT Centennial: “If computers of the kind I have advocated become the computers of the future, then computing may someday be organized as a public utility just as the telephone system is a public utility... The computer utility could become the basis of a new and important industry.” \cite{Sadashiv2011} His concept was the basis for the idea of Douglas Parkhill \cite{Hannabuss2009, Hashemi2012, Sadashiv2011, Vandenberg2005} to create a grid computing paradigm that was described later in 1966 and was a set of computers connected over a grid that take the computing decisions collectively.

The fog computing approach was one of the first technologies to solve the latency issues of cloud computing. The "Fog Computing" term was first proposed by CISCO in 2012 \cite{Haouari2018} and has been described as "a highly virtualized platform that provides compute storage, and networking services between end devices and traditional Cloud Computing Data Centers, typically, but not exclusively located at the edge of the network" \cite{Bonomi2012}. The OpenFog group was established in 2015 to develop standards in the field of fog computing. It included companies and academic organizations such as Cisco, Dell, Intel, Microsoft Corp, and Princeton University. On December 18, 2018, the OpenFog consortium became part of The Industrial Internet Consortium \cite{IndustrialInternetConsortium}. 

\tab{tab-timeline}{Fog computing timeline}{
\resizebox{\textwidth}{!}{
\begin{tabular}{ |c|c|c|c|c| } 
 \hline
 1961 & 1990's & 2012 & 2015 & 2018 \\ 
 \hline \hline
 \makecell[l]{\textbf{John McCarthy}.\\ Utility computing \\Definition \cite{Russo2020}} 
 & \makecell[l]{\textbf{Ian Foster et. al}.\\Definition of\\ the grid computing \cite{Foster2011}}
 & \makecell[l]{\textbf{Flavio Bonomi et. al}.\\CISCO proposed\\ the definition\\ of the cloud computing \cite{Bonomi2012}}
 & \makecell[l]{\textbf{The OpenFog group}\\ was established \cite{OpenFogConsortiumArchitectureWorkingGroup2017}}
 & \makecell[l]{\textbf{Machaela Iorga et. al}.\\ The NIST published\\ the definition\\ of the fog computing \cite{Iorga2018}}
 \\
 & 
 & \makecell[l]{\textbf{Mell Peter}.\\ The NIST published\\ the definition\\ of the cloud computing \cite{Mell2012}}
 & 
 & \makecell[l]{\textbf{Mahmoudi Charid}.\\ The formal definition\\ of the edge computing\\ was published \cite{MahmoudiMourlinBattou2018}}
\end{tabular}
}
}

In 2018, the National Institute of Standards and Technology of the United States had formulated an official definition of the fog computing term: "Fog computing is a layered model for enabling ubiquitous access to a shared continuum of scalable computing resources. The model facilitates the deployment of distributed, latency-aware applications and services, and consists of fog nodes (physical or virtual), residing between smart end-devices and centralized (cloud) services. The fog nodes are context-aware and support a common data management and communication system. They can be organized in clusters - either vertically (to support isolation), horizontally (to support federation), or relative to fog nodes’ latency-distance to the smart end-devices. Fog computing minimizes the request-response time from/to supported applications, and provides, for the end-devices, local computing resources and, when needed, network connectivity to centralized services" \cite{Iorga2018}. 

Bridging the gap between the cloud and end devices through computing, storage, and data management not only in the cloud but also on intermediate nodes \cite{Liu2017} has expanded the scope of fog computing, which allowed its application in new tasks such as IoT, smart vehicles \cite{Huang2017}, smart cities \cite{Dantas2019}, health care \cite{Gu2017}, smart delivery (including the use of drones) \cite{Wadhwa2019}, video surveillance, etc. \cite{Zhang2018}. These systems benefit significantly from Big Data processing \cite{Russo2020}, allowing them to extract new knowledge and decision-making information from the data streams generated by clusters of IoT devices. Fog computing supports this challenge by enabling distributed computing resources for lightweight data processing tasks, including filtering and preprocessing data before sending it to the cloud. But the geographical distribution, heterogeneity of computing nodes, and high instability of network communications at the edge level lead to the need to solve complex problems associated with monitoring, scheduling, and ensuring the necessary quality of service of such services.

\subsection{Key characteristics of fog computing} \label{Key characteristics of fog computing}
Due to the late separation of the fog and edge computing concepts, many companies introduced their characteristics \cite{Al-Doghman2017} and definitions for fog and edge computing, often combining them into one \cite{Liu2017}. \tabref{tab-characteristics} presents the key characteristics that different authors distinguished for fog and edge computing. 

In 2017, the OpenFog Consortium released a reference architecture for fog computing, which is based on eight basic principles: security, scalability, openness, autonomy, RAS (reliability, availability, and serviceability), agility, hierarchy, and programmability \cite{OpenFogConsortiumArchitectureWorkingGroup2017}. 

\tab{tab-characteristics}{Characteristics of Fog Computing \cite{Naha2018}}{
\centering
\resizebox{\textwidth}{!}{\begin{tabular}{|l|l|l|l|l|l|l|l|l|l|l|} 
\hline
\multirow{2}{*}{\textbf{ Method }} & \multicolumn{10}{l|}{Properties} \\ 
\cline{2-11}
 & \begin{sideways}\textbf{ Highly virtualized  }\end{sideways} & \begin{sideways}\textbf{ Generally used for IoT }\end{sideways} & \begin{sideways}\textbf{ Extends the cloud }\end{sideways} & \begin{sideways}\textbf{ Not exclusively located at the edge }\end{sideways} & \begin{sideways}\textbf{ Resides at network ends  }\end{sideways} & \begin{sideways}\textbf{ \makecell{Fog device consists of processing,\\ storage, and network connectivity} }\end{sideways} & \begin{sideways}\textbf{ Run in a sandboxed environment }\end{sideways} & \begin{sideways}\textbf{ \makecell{Leasing part of users \\ devices and provide an incentive} }\end{sideways} & \begin{sideways}\textbf{ Fog and Edge computing as similar }\end{sideways} & \begin{sideways}\textbf{ Can be deployed anywhere }\end{sideways} \\ 
\hline
\textbf{ Bonomi et al. [68] } & + & + & ~ & + & ~ & ~ & ~ & ~ & ~ & ~ \\ 
\hline
\textbf{ Cisco Systems [37] } & ~ & + & + & ~ & ~ & + & ~ & ~ & ~ & + \\ 
\hline
\textbf{ Vaquero and~  Rodero-Merino [86] } & ~ & ~ & ~ & ~ & ~ & + & + & + & ~ & ~ \\ 
\hline
\textbf{ IBM [3] } & ~ & ~ & + & + & + & ~ & ~ & ~ & + & ~ \\ 
\hline
\textbf{ Synthesis [51] } & + & ~ & + & + & + & ~ & ~ & ~ & ~ & ~ \\
\hline
\end{tabular}}
}

In \cite{Hong2019} and \cite{Anawar2018}, the following key characteristics of fog computing are highlighted. 
\begin{itemize}
    \item \textit{Contextual location awareness and low latency. }Fog computing offers the lowest-possible latency due to the fog nodes’ awareness of their logical location in the context of the entire system and of the latency costs for communicating with other nodes. 
    \item \textit{Geographical distribution.} In sharp contrast to the more centralized cloud, the services and applications targeted by fog computing demand widely but geographically identifiable, distributed deployments.
    \item \textit{Heterogeneity.} Fog computing supports the collection and processing of data of different form factors acquired through multiple types of network communication capabilities 
    \item \textit{Interoperability and federation.} Seamless support of certain services (real-time streaming services is a good example) requires the cooperation of different providers. Hence, fog computing components must be able to interoperate, and services must be federated across domains. 
    \item \textit{Real-time interactions.} Fog computing applications involve real-time interactions rather than batch processing. 
    \item \textit{Scalability and agility of federated, fog-node clusters.} Fog computing is adaptive, at cluster or cluster-of-clusters level, supporting elastic compute, resource pooling, data-load changes, and network condition variations, to list a few of the supported adaptive functions.
    \item \textit{Cognition. }Cognition is responsiveness to client-centric objectives. Fog-based data access and analytics give a better alert about customer requirements, best position handling for transmitting, storing, and controlling functions throughout the cloud to the IoT continuum. Applications, due to proximity, at end devices provide a better conscious and responsive reproduced customer requirement relation \cite{Yin2015}.
    \item \textit{Support for Mobility.} Mobility support is a vital fog computational advantage that can enable direct communication between mobile devices using SDN protocols (i.e., CISCO Locator/ID Separation Protocol) that decouples host identity from location identity with a dispersed indexing system \cite{Zhu2013}.
    \item \textit{Large Scale Sensor Network.} The fog has a feature applicable when an environment monitoring system, in near smart grid applications, inherently extends its monitoring systems caused by hierarchical computing and storage resource requirements.
    \item \textit{Widespread Wireless Access.} In this scenario, wireless access protocols (WAP) and cellular mobile gateways can act as typical examples of fog node proximity to the end-users.
    \item \textit{Interoperable Technology.} Fog components must work in an interoperating environment to guarantee support for a wide range of services like data streaming and real-time processing for best data analyses and predictive decisions.
\end{itemize}

\subsection{Fog and Edge Computing Concepts Definitions} \label{Fog and Edge Computing Concepts Definitions}
Some sources refer to fog computing as edge computing, relying on the critical technology feature that data collection and analysis is not organized in a centralized cloud, but as close to the end device as possible, "at the edge of the network" \cite{Bonomi2012, Garcia2019, Hong2017, Iorga2018}. 

However, \cite{Yousefpour2019} indicates that although fog and edge computing move computation and data storage closer to the network edge, these paradigms are not identical. 
Within the Fog Computing paradigm, fog nodes are located at the edge of the local network, often they are deployed based on routers and wireless access points (if these devices support the required technologies for deployment of the fog node) \cite{Wadhwa2019}. In contrast to fog computing, edge computing is deployed even "closer" to the end devices, already inside the local network itself on the intermediate access points. Sometimes the end devices themselves can act as edge computing nodes. Smartphones, tablets, and other computing devices with sufficient computing capabilities and support for the deployment of computing nodes can handle edge computing tasks \cite{Tuli2019}. However, this also limits their computational power, and therefore there are some limitations in their application scope. So edge computing is used to solve such tasks as video surveillance, video caching, and traffic control \cite{Yousefpour2019}.

The OpenFog Consortium claims that edge computing is often erroneously referred to as fog computing and determine that the main difference is fog computing is the overall architecture of distributing resources across the network, whereas edge computing is specifically focused on executing compute processes close to end-users outside the core of the network \cite{MahmoudiMourlinBattou2018}. In \cite{Chiang2017}, the authors note on fog and edge computing that "fog is inclusive of cloud, core, metro, edge, clients, and things" and "the fog seeks to realize a seamless continuum of computing services from the cloud to the things rather than treating the network edges as isolated computing platforms”. 

Thus, the term "edge computing" is mainly used in the telecommunications industry and usually refers to 4G/5G, RAN (Radio Access Network), and ISP (Internet Service Provider) base stations \cite{Chiang2017, Li2018}. However, this term has recently been used in the subject area of IoT \cite{GEDigital2018, Li2018, AGuideToEdgeIoTAnalytics} concerning the local network where sensors and IoT devices are located. In other words, "edge computing" is located within the first of the IoT device of the transit section of the network, for example, at WiFi access points or gateways.

\subsection{Classification of fog computing applications} \label{Classification of fog computing applications}
Fog computing enables new applications, especially those with strict latency constraints and those involving mobility. These new applications will have heterogeneous QoS requirements and demand Fog management mechanisms to cope efficiently with that heterogeneity. Thus, resource management in Fog computing is quite challenging, calling for integrated mechanisms capable of dynamically adapting the allocation of resources. The very first step in resource management is to separate the incoming flow of requests into Classes of Service (CoS) according to their QoS requirements. The mapping of applications into a set of classes of service is the first step in creating a resource management system capable of coping with the heterogeneity of Fog applications. The authors of \cite{Guevara2020} proposed the following critical classes of fog computing applications:
\begin{itemize}
    \item \textit{Mission-critical.} Applications in which a component failure would cause a significant increase in the safety risk for people and the environment. Those are healthcare systems, criminal justice, drone operations, industrial control, financial transactions, military, and emergency operations. Those applications should implement distribution features to ensure duplication of functionality.
    \item \textit{Real-time.} The speed of response in these applications is critical since data are processed at the same time they are generated but can tolerate a certain amount of data loss (online gaming, virtual and augmented reality applications).
    \item \textit{Interactive.} Responsiveness is critical; the time between when the user requests and actions is less than a few seconds.  Those are interactive television, web browsing, database retrieval, server access applications.
    \item \textit{Conversational.} Characterized by being delay-sensitive but loss-tolerant with slight delays (about 100-200 ms). E.g., video and Voice-over-IP (VoIP) applications where losses cause occasional glitches in audio or video playback.
    \item \textit{Streaming} class applications are accessed by users on-demand and must guarantee interactivity and continuous playout. The network must provide each stream with an average throughput that is larger than the content consumption rate. In such a case, data should be located as close to the end-user as possible, and new nodes should easily be created and removed from the environment.
    \item \textit{CPU-bound.} Involves complex processing models, such as those in decision making, which may demand hours, days, or even months of processing. Face recognition, animation rendering, speech processing, and distributed camera networks are examples of this applications class.
    \item \textit{Best-effort.} For these applications, long delays are annoying but not particularly harmful; however, the completeness and integrity of the transferred data are of paramount importance. Some examples of the Best-Effort class are e-mail downloads, chats, SMS delivery, FTP, P2P file sharing.
\end{itemize}

\section{Technologies that support fog and edge computing} \label{Technologies that support fog and edge computing}
\subsection{Virtualization} \label{Virtualization}
 The key technology that supports cloud and fog computing is virtualization \cite{Tseng2018}, which allows you to use the resources of one physical machine by several logical virtual machines (VMs) at the level of Hardware Abstraction Layer (HAL). Virtualization technology uses a hypervisor - a software layer that provides the operation of virtual machines based on hardware resources. A machine with a hypervisor is called a host machine. A virtual machine running on the host machine is called a guest machine, on which in turn the guest operating systems (OS) can be installed. This type of virtualization is called hypervisor-based virtualization.

There is also container-based virtualization \cite{DeDonno2019}, representing a packaged, standalone, deployable set of application components that can also include middleware and business logic in binary files and libraries to run applications. 

Authors of \cite{Radchenko2019} present a comparative analysis of both types of virtualization, based on which we can highlight some of the advantages of container-based virtualization. 
\begin{itemize}
    \item \textit{Hardware costs. }Virtualization via containers decreases hardware costs by enabling consolidation. It enables concurrent software to take advantage of the true concurrency provided by a multicore hardware architecture.
    \item \textit{Scalability. }A single container engine can efficiently manage large numbers of containers, enabling additional containers to be created as needed. 
    \item \textit{Spatial isolation.} Containers support lightweight spatial isolation by providing each container with its resources (e.g., core processing unit, memory, and network access) and container-specific namespaces. 
    \item \textit{Storage.} Compared with virtual machines, containers are lightweight concerning storage size. The applications within containers share both binaries and libraries. 
    \item \textit{Real-time applications.} Containers provide more consistent timing than virtual machines, although this advantage is lost when using hybrid virtualization. 
    \item \textit{Portability.} Containers support portability from development to production environments, especially for cloud-based applications.
\end{itemize}

Thus, two main virtualization technologies are currently used to support fog computing \cite{Kakakhel2018}: hypervisor-based and container-based. Cloud computing mainly uses hypervisor-based virtualization to share limited hardware resources between several virtual machines. Fog computing that commonly hosted on low-performance hardware prefers container-based virtualization to create node instances on new hardware devices. That’s why container-based virtualization is becoming more and more widespread in fog computing. Due to lower hardware performance requirements to ensure the deployment of computing nodes, intermediate devices may not have high computing power. This is especially relevant for edge computing nodes because they are not even run on the IoT devices themselves \cite{Puliafito2018} but on intermediate access points closest to the IoT devices.

\subsection{Fog computing orchestration} \label{Fog computing orchestration}
With containerization evolving as one of the technologies to support fog computing, the challenge arose to manage the computational load to ensure efficient use of geographically dispersed resources \cite{Jiang2018}. Fog computing implementation requires a different level of computing resource management compared to the cloud, for example \cite{Velasquez2018}. 

The first complex task that arises when working with fog computing, as opposed to cloud computing, is managing the distribution of computational load (orchestration) between nodes of the fog \cite{Li2018, Liu2005} by placing the fog services on them, as well as orchestration of these services, i.e. ensuring efficient collaboration of computational services for solving tasks assigned to the fog environment. 
Authors of \cite{Wen2017} formulate that orchestration provides the centralized arrangement of the resource pool, mapping applications with specific requests and providing an automated workflow to physical resources (deployment and scheduling); workload execution management with runtime QoS control; and time-efficient directive generation to manipulate specific objects.

Let us consider the key tasks to be solved by the Fog Orchestrator \cite{DeBrito2017a, Velasquez2018}. 
\begin{itemize}
    \item \textit{Scheduling.} It is necessary to consider how to exploit the collaboration between nodes to offload applications efficiently in Fog environments. In general, the processing nodes should be managed by a resource broker in the Orchestrator to perform smart scheduling of the resource, considering the applications’ workflows. 
    \item \textit{Path computation's} main objectives are: maintaining end-to-end connectivity, adapting to dynamic topologies, maximizing network and application traffic performance and providing network resilience. 
    \item \textit{Discovery and allocation} of the physical and virtual devices in the Fog, as well as the resources associated with them. 
    \item \textit{Interoperability} is the ability that distributed system elements are able to interact with each other. Several factors influence the interoperability of a system, such as the heterogeneity of its elements.
    \item \textit{Latency. }One of the characteristics of Fog environments is that they provide low levels of latency. This allows the deployment of a different kind of services with real-time and low latency restrictions that are not necessarily fit for the cloud; but also requires a new set of mechanisms that guarantee that these low latency levels are met. 
    \item \textit{Resilience. }To guarantee a smooth work of the complex and diverse environment where the IoT acts from the resilience perspective, an Orchestrator should be in charge of intelligent migration and instantiation of resources and services providing a global view of the status of the IoT. 
    \item \textit{Prediction and optimization.} Proper management of resources and services in an IoT environment, where these are geographically distributed, generating multi-dimensional data in enormous quantities, is only possible if the orchestration process takes into consideration prediction and optimization mechanisms of all overlapping and interconnected layers in the IoT. 
    \item \textit{Security and privacy.} From the privacy perspective, the main challenge lies in preserving the end-user privacy since the Fog nodes are deployed near them, collecting sensitive data concerning identity and usage patterns. Regarding security, a significant challenge is how to deal with the massively distributed approach of the Fog to guarantee the proper authentication mechanisms and avoid massive distributed attacks. 
    \item \textit{Authentication, access, and account.} To perform activities related to application life cycle management (i.e. deployment, migration, application of policies), the Orchestrator interacts with the fog nodes in the environment.
\end{itemize}

Optimization of various metrics (latency, bandwidth, energy consumption etc.) plays a vital role in fog computing orchestration. The following key tasks related to the distribution of tasks and data by the level of fog computing are currently being identified \cite{Bellendorf2020}:
\begin{itemize}
    \item Offloading computing tasks from end devices to fog nodes and cloud.
    \item Scheduling of tasks within a fog node.
    \item Clustering of fog nodes: how to determine the size of a cluster of fog nodes to handle the relevant requests
    \item Migration of data/applications between fog nodes.
    \item Geographical distribution of physical resources (before operation).
    \item Distributing applications/data among fog nodes and cloud.
\end{itemize}

\subsection{Fog computing and security issues} \label{Fog computing and security issues}
Due to the significant degree of decentralization of the computing process, security in fog computing differs in some critical aspects from mechanisms used, for example, in cloud computing. The design of a secure fog architecture must take into account the security features of each layer of the computing architecture, including the features of lightweight wireless data transfer at the sensing/edge layer; data transfer over middleware mesh networks; preprocessing of data using clusters of fog nodes on the application level; possible data transfer over the WAN for processing in the public cloud \cite{Aljumah2018, Puthal2019}.

Each of these layers has its security issues and vulnerabilities. The sensing layer is vulnerable to sensors and devices being targets of outcoming threats, including device tampering, spoofing attacks, signal attacks, malicious data, etc. At the middleware level, the secure transmission of sensed data and its storage are the primary concerns. This layer deals with confidentiality, integrity, and availability issues. The security requirements at the application layer are determined directly by the application being executed. \figref{pic/Fig-Security} presents a classification of possible security issues and their solutions for each of the fog architecture layers listed above \cite{Aljumah2018}. 

The authors of \cite{Yakubu2019} state that the most promising research directions for security solutions in fog computing are cryptographic techniques and machine-learning for intrusion detection. Cryptographic processing includes encryption, decryption, key and hash generation, and verification of hashes used to guarantee data privacy.

As an example of this technique, the Configurable Reliable Distributed Data Storage System \cite{Chervyakov2019} was designed to secure data flows in whole fog. Such a system uses the AR-RRNS (Approximation of the Rank - Redundant Residue Number System) method to encrypt and decrypt data using error correction codes and secret sharing schemes. Machine-learning techniques are proposed to analyze data flow and node states to detect outside intrusion. To implement such a traffic analysis, the fog orchestrator can act as a tool to detect an intrusion or data corruption \cite{Fakude2019}.

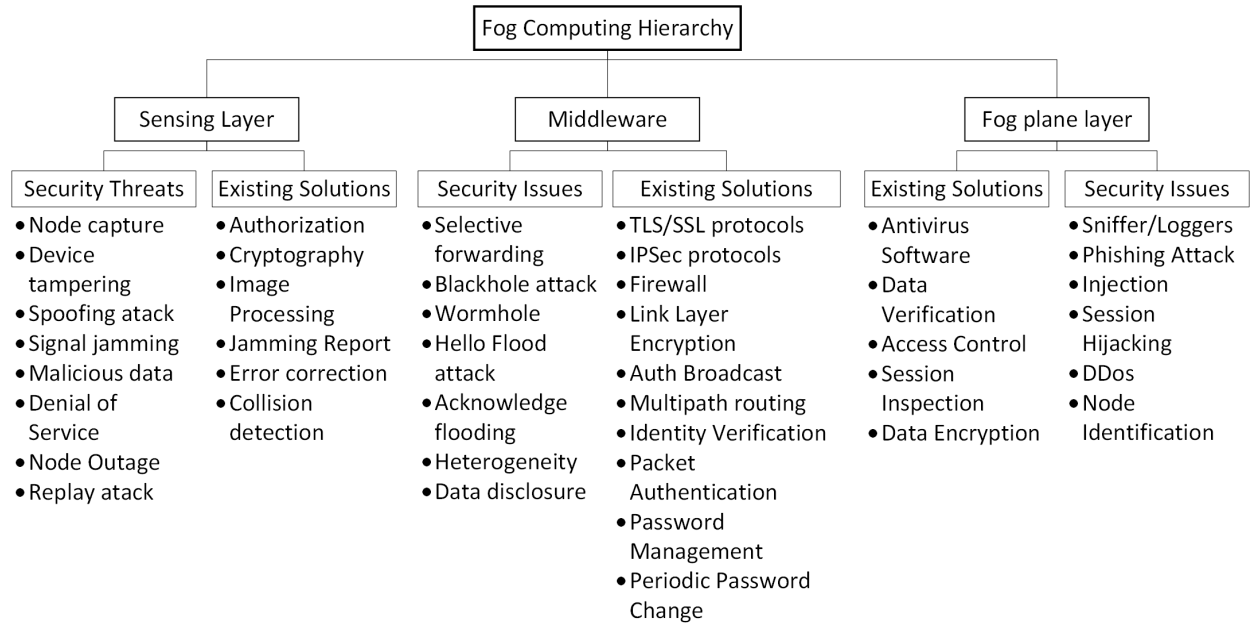
\fig{width=1\textwidth}{pic/Fig-Security}{The security threats and solutions classifications in fog computing. DDoS: distributed DoS; TLS: transport layer security; SSL: secure sockets layer; IPsec: Internet Protocol security \cite{Aljumah2018}}

\subsection{Fog computing and scalability} \label{Fog computing and scalability}
Scalability is another essential feature for fog computing systems to adapt workload, system cost, performance, and business needs. Based on fog computing hierarchical properties, we can highlight the following key elements of fog architecture that can be scaled \cite{Tran2020}:
\begin{itemize}
    \item \textit{Virtual Nodes:} through software reconfiguration, specifying if several virtual nodes can be placed on one physical device;
    \item \textit{Physical Nodes: }vertical scalability trough hardware upgrade;
    \item \textit{Networks:} horizontal scaling of fog nodes and adapting to environmental changes and dynamic workloads.
\end{itemize}

Adding new fog nodes to the fog network affects all three main aspects of scalability discussed above. However, this task commonly requires manual workload from network administrators, while it is hard to effectively identify the location or cluster of the new fog node. In \cite{Tran2020} the fog model that helps to overcome this difficulty was proposed. It automates the introduction of new fog nodes into an existing network based on the current network conditions and services required by customers. Concretely, the newly added fog node can detect its geographical location (e.g., using its network scan capability or via its GPS module) and identify the most suitable cluster to connect with.

Kubernetes platform is now the de-facto standard for service management in centralized distributed systems such as clouds. In this regard, its application to the management of fog infrastructures is of definite scientific interest. Kubernetes has a native mechanism for auto-scaling that considers only CPU usage. Users can specify the maximal number of application instances, but the actual number of application instances activated is under the control of Kubernetes. Authors of \cite{Zheng2019} developed a modification of the scheduling algorithm based on the Kubernetes platform to manage resource autoscaling tasks in fog computing systems. A fog computing platform was designed as a collection of physically distributed containers that are orchestrated by Kubernetes and AS Broker – a service running in a Pod on Master. It communicates through APIServer with the Controller and Scheduler of Kubernetes to obtain a list of nodes where application instances are currently running. It then collects node information from all nodes. If the number of application instances should be adjusted, it sends a request through APIServer for Pod number adjustment. 

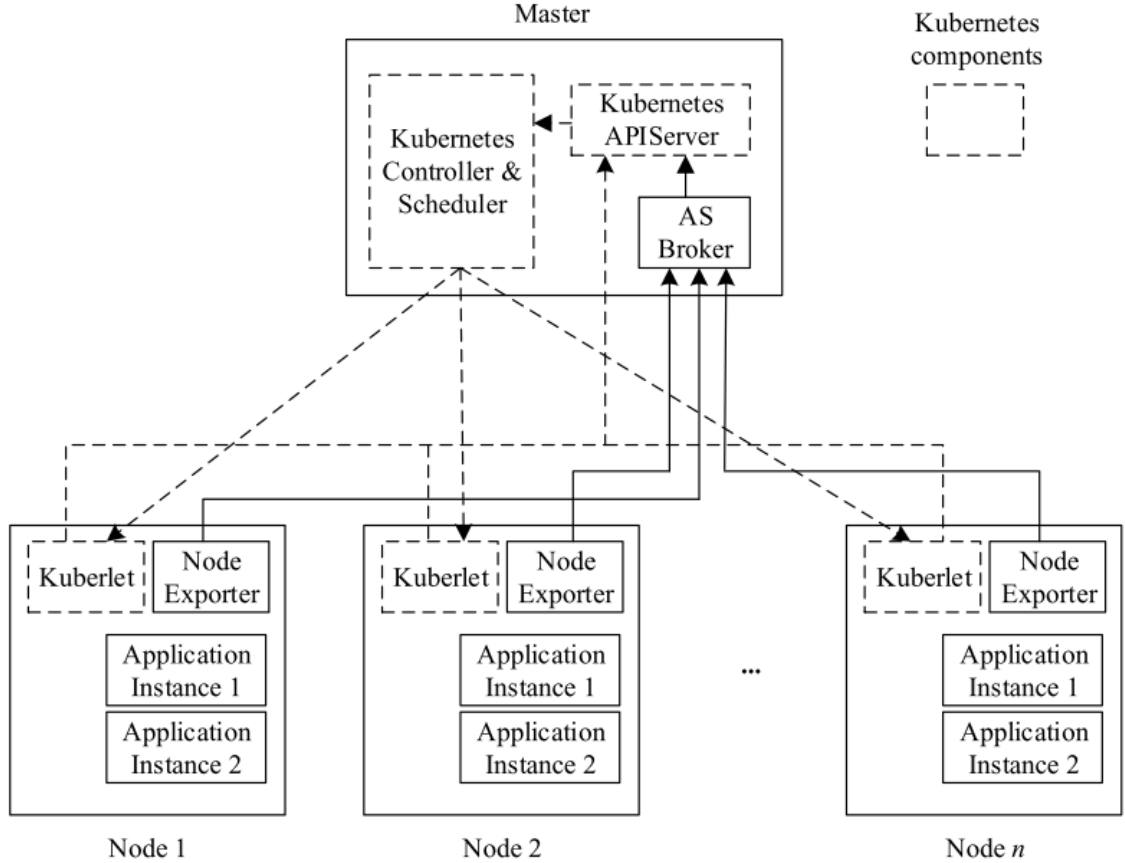
\fig{width=1\textwidth}{pic/Fig-Proposed-Architecture}{Proposed architecture of fog network based on Kubernetes \cite{Zheng2019}.}

The scalability experiment in \cite{Zheng2019} included four independent fog nodes. A stress program was used to generate CPU and memory load to emulate the processing of requests. Every request took a 15-second execution time and a 50 MB memory amount. \figref{pic/Fig-Application-Response} shows tested application response time with and without AS Broker. Though response time dynamically changes, the result with AS Broker is better than that without almost at every time point. This result demonstrates the effectiveness of the proposed scheme.
 
\fig{width=1\textwidth}{pic/Fig-Application-Response}{Application response time with and without AS Broker \cite{Zheng2019}}

The authors of \cite{Fahs2020} also investigate the possibilities of automatic scaling of computing resources of fog platforms. The objective of their work was to dynamically scale and place an application’s replicas in a cluster of geo-distributed fog nodes to predominantly minimize the number of slow requests while maintaining efficient resource utilization. As a critical parameter determining the quality of service, the authors use the average response time parameter whether the latency and the processing capacity requirements are still met. The following methods were used to improve the quality of service: transferring service from one node to another, load redirection to the nearest lightly loaded servers, or scaling, by creating new replicas of computing services. The experimental setup consisted of 22 Raspberry Pi (RPi) model 3B+ single-board computers acting as fog computing servers. The RPis were organized with one master node and 21 worker nodes capable of hosting replicas. \figref{pic/Fig-Evaluation} shows the average number of placements that could be studied per second for various system sizes with the average number of placements that had to be evaluated to repair a latency or capacity violation. When the cluster size increased, the time needed to study any single placement also increased. However, even for a large system with 500 nodes, Voil`a evaluated approximately 100 placements per second. 
 
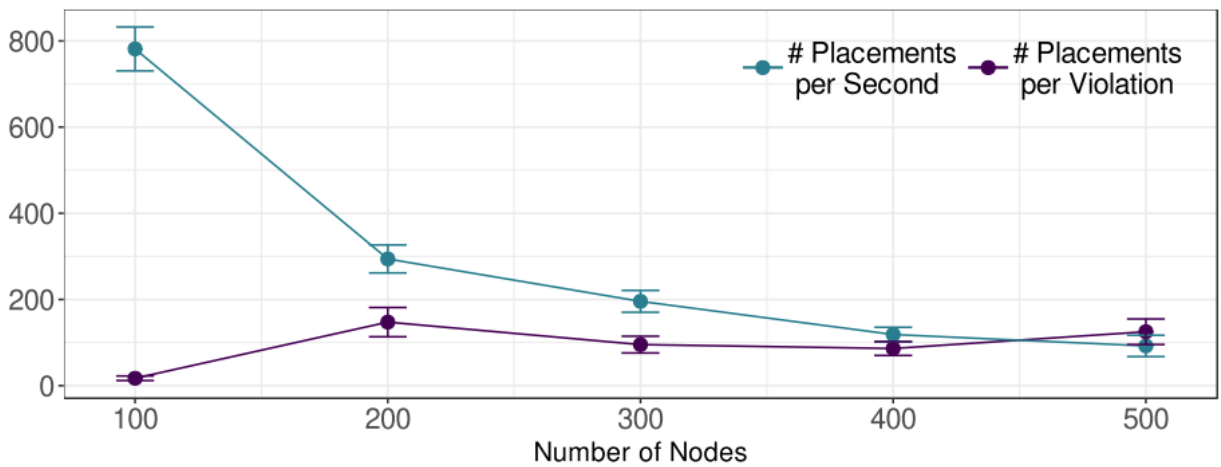
\fig{width=1\textwidth}{pic/Fig-Evaluation}{Evalutation of average number of placement that can be studied per second \cite{Fahs2020}}.

The idea of using fog computing as a computing swarm and architecture of organizing fog nodes and application was described in \cite{Brogi2018}. In the proposed architecture, the fog consists of fog nodes (FNs), fog node controllers (FNCs), applications (Apps) and application controllers (ACs). FNCs control when FNs can be attached and detached.  With the help of ACs FNCs can scale up and down the set of resources assigned to an App (e.g. by decreasing/increasing the cores, CPU time, and bandwidth assigned to such App) by simply changing the resources assigned to the corresponding Docker container. If the computational resources available in a FN are no more capable of satisfying the requirements of all Apps running on it, the FNC of the overloaded FN will interact with the other FNCs in the system to decide which Apps can be migrated and on which FNs.

Attempts to use fog computing with low-power devices to solve resource-intensive computational problems have been made since the beginning of the concept of fog computing. Authors of \cite{Tsai2017}  use the resources of low-power Raspberry Pi-based nodes for Machine Learning Data Analytics using Deep Learning. They took the SCALE (Safe Community and Alerting Network) TensorFlow Application that uses various sensors (movement detection, gas, temperature, heartbeat, etc.) to build a security system as an example. The authors enhanced this project to support two crucial sensors: camera and microphone. They collected sensor data along the path where the person with the camera passed. \figref{pic/Fig-Different-Service-Quality}(a) shows the number of processed images per minute when running enhanced application on two devices with eight different cutting points. The application implemented a 9-layers network, so cuts are made between layers. When the cutting point went from 1 to 8, more complicated operators were put on the first device. The first device processed images before the second device. As shown in \figref{pic/Fig-Different-Service-Quality}(a), cutting points 4 and 5 resulted in the best performance. It is explained by \figref{pic/Fig-Different-Service-Quality}(b), which shows that cutting an application into smaller operators with similar complexity results in the best performance. Moreover, \figref{pic/Fig-Different-Service-Quality}(c) reports the network overhead caused by distributed analytics. It shows that if more loads were put on the first device, it resulted in lower network overhead. Hence, when network resources were the bottleneck, equally-loaded splitting decisions were not preferred.
 
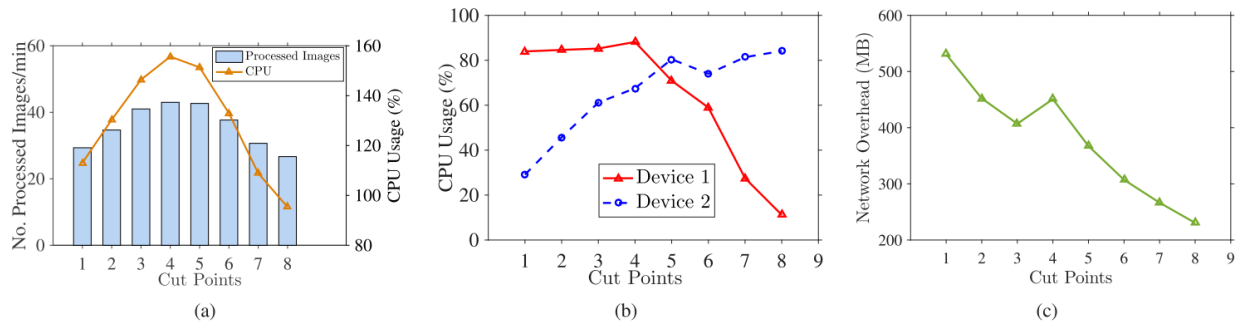
\fig{width=1\textwidth}{pic/Fig-Different-Service-Quality}{Different service quality caused by different cutting points: (a) the number of processed images, (b) the CPU and RAM usages, and (c) the network overhead  \cite{Tsai2017}.}

\section{Overview of Fog Computing Platforms} \label{Overview of Fog Computing Platforms}
While reviewing existing fog computing deployment platforms, we would consider commercial as well as open-source platforms. The complexity of the analysis of commercial platforms is the lack of information about their architecture and the technical solutions used, which constitute a trade secret. However, the analysis of commercial solutions has shown that among commercial fog platforms, there are platforms with the full support of fog computing (computing, analytics, and organization of the transport layer of the fog network) and platforms that provide only the transport layer of the fog network and do not provide management of computing nodes and fog computing itself. Platforms that provide only the transport layer of fog computing will not be considered in this paper. 

The following key characteristics of private and public commercial fog platforms can be highlighted (see \tabref{tab-private-fog} and \tabref{tab-public-fog}).
\begin{itemize}
    \item \textit{Supported hardware platforms} - the platform can work with any device that supports virtualization or containerization, or only with a limited list of devices - through drivers or branded devices. Smartly Fog, ThingWorx, and Cisco IOx only work with their proprietary hardware. 
    \item \textit{Basic development technology} - which executable environment is used to create, deploy and run fog applications.
    \item \textit{Open communication protocols and SDK} - is there any restriction on the applications that can be used in the fog: whether it is necessary to port applications, or in principle can be executed only applications written using special supplied SDK, as in the case of ThingWorx, whose fog applications should be written using a proprietary SDK to run in the fog. 
    \item \textit{Deployment technology} - which of the technologies is used to deploy fog nodes, if known. \item Integration options - is it possible to integrate with other platforms, such as enterprise solutions or public clouds? 
    \item \textit{Connecting of external data sources} - the platform's ability to connect to third-party databases and data warehouses physically located outside the central cloud for data storage and processing. 
    \item \textit{Availability of additional services (Machine Learning, Analytics, etc.)} - the ability to connect and use additional services, which provide additional functionality for analysis and work with data in the fog. 
    \item \textit{Edge support} - the ability to connect and use edge devices and edge computing, and further collect and process information from them.
\end{itemize}

\subsection{Private fog platforms} \label{Private fog platforms}
Private fog platforms provide private fog solutions based on computing infrastructure deployed directly on the customer's resources.
\tab{tab-private-fog}{Overview of private fog platforms}{
\centering
\resizebox{\textwidth}{!}{\begin{tabular}{|l|c|c|c|c|c|c|} 
\hline
\multicolumn{1}{|c|}{Feature} & \textbf{ClearBlade} & \textbf{\makecell{Smartiply\\Fog}} & \textbf{LoopEdge} & \textbf{ThingWorx} & \textbf{\makecell{Nebbiolo\\Technologies}} & \textbf{\makecell{Cisco\\IOx}} \\ 
\hline
\textbf{\makecell[l]{Supported\\hardware platforms}} & Universal & Own equipment & Universal & Own equipment & Universal & Own equipment \\ 
\hline
\textbf{\makecell[l]{Basic\\development technology}} & JavaScript & No data & Universal
  (Docker) & Java VM & Universal
  (Docker) & \makecell{Docker,\\Linux,\\IOx} \\ 
\hline
\textbf{\makecell[l]{Open communication\\protocols and SDK}} & + & + & + & – & + & + \\ 
\hline
\textbf{Deployment technology} & Linux KVM & No data & Docker & No data & Docker & Linux KVM \\ 
\hline
\textbf{Integration opportunities} & \makecell{Oracle,\\SAP,\\Microsoft,\\Salesforce} & – & – & \makecell{Microsoft,\\Azure IoT,\\Hub} & – & \makecell{Microsoft,\\Azure IoT,\\Hub} \\ 
\hline
\textbf{\makecell[l]{Connecting external\\data sources}} & + & – & + & + & + & + \\ 
\hline
\textbf{\makecell[l]{Availability of\\additional services}} & No data. & + & – & + & + & + \\ 
\hline
\textbf{ Edge Support } & + & + & + & + & + & + \\
\hline
\end{tabular}}
}

\textbf{Cisco IOx platform} was presented by Cisco in 2014 \cite{Antonio2014} as a network infrastructure development due to the expected growth of IoT. The platform's focus is to reduce the labor costs of porting applications to the fog nodes, achieved through containerization technologies and based on its operating system based on the Linux OS.

Cisco IOx is an application environment that combines Cisco IOS (a mini operating system of Cisco hardware) and Linux. Open-source Linux utilities are used to develop applications. It uses a single protocol for the interaction of fog applications throughout the network, organized using Cisco IoT technologies. Both Cisco and its partners supply IOx infrastructure fog applications. A variety of general-purpose programming languages is supported to develop Cisco IOx applications.

Docker is used for deploying applications. Various types of applications are supported, including Docker containers and virtual machines (if network equipment has such a capability). It is also possible to use your IOx executable environment to write applications in high-level programming languages (such as Python). 

\textbf{The Nebbiolo Technologies platform} is aimed at the corporate industrial market, supporting the Industry 4.0 concept \cite{NebbioloTechnologiesInc.2016}. Nebbiolo Technologies closely cooperates with Toshiba Digital Solutions \cite{ToshibaDigitalSolutions} in supplying complete computing solutions for the industrial and IoT sectors. 

The platform consists of fogNode hardware, fogOS software stack, and fogSM system administrator, deployed in the cloud or locally \cite{Hong2019}. Fog System Manager (fogSM) provides a cloud-based, centralized management platform that allows you to deploy and configure devices on the periphery. 

The platform's key feature is fogOS \cite{Hong2019} - a software stack that provides communication, data management and application deployment at the fog level. Based on a hypervisor, fogOS provides a set of functions in a virtualized form. It supports a wide range of device connectivity standards and allows applications to be hosted and managed in real-time. 

\textbf{ClearBlade Platform} is a technology stack that provides fast development and deployment of enterprise IoT solutions, from edge devices to cloud services. It includes software components installed on the entire IoT device stack and the ability to connect third-party systems through the provided API for integration with devices, internal business applications, and cloud services. ClearBlade Platform provides a centralized console for managing IoT applications, with the ability to deploy locally or in the cloud. Platform management functions are delegated to the edge nodes (or on the end devices themselves or their gateways) using ClearBlade Edge fog and edge computing \cite{Hughes2017}.

The platform supports a serverless computing approach to the development of services based on the JavaScript language, which can be configured to implement machine learning and data analysis methods. The platform provides mechanisms for exporting data and analytics collected by the system to widely used business systems, applications, and databases through integration with corporate platform solutions from Oracle, SAP, Microsoft, and Salesforce. ClearBlade also provides in-house dashboards, business applications, and database management systems for integrated monitoring and management of the IoT ecosystem.

ClearBlade uses the OAuth model for access control, where each user and device receives a token that must be authorized to gain access to the system or its node. The data is encrypted on the devices themselves as well as on network transmissions. Transmitted data is encrypted using OpenSSL libraries with TLS encryption. 

\textbf{The Smartiply Fog platform} is a cloud computing platform that focuses on optimizing resources and keeping your devices running even without connecting to the cloud. The platform provides greater reliability for online environments by optimizing resources and computing based on proprietary hardware \cite{EdgeGatewaySmartiply}. The platform enables point-to-point interaction between devices. In this way, the node system can continue to operate autonomously to receive, analyze and store data, up to restoring communication with the external network \cite{MobilePlatformSmartiply}. 

\textbf{LoopEdge platform} from Litmus Automation \cite{LitmusRef, LoopEdgeRef} allows you to connect different devices in a single system, collect and analyze data from them. Litmus Automation also provides a Loop platform that allows you to manage the life cycle of any IoT device and export real-time data to internal analytical and business applications. This platform is widely distributed among well-known engineering concerns: Nissan, Renault, Mitsubishi Corporation Techno. 

The  platform developers emphasize that it can work with virtually any device, industrial and domestic consumers. For example, the platform supports the connection of devices based on Arduino and Raspberry Pi. Even if some device is not supported, connecting it to the platform is relatively easy due to the executable packages installed on the device itself, which can be expanded and created from scratch for a particular device.

\textbf{PTC ThingWorx} platform \cite{ThingworxRef} is an IoT platform that offers the connection possibility to more than 150 types of devices. However, since devices are connected through drivers that require installation, this platform is not universal and has limitations on the devices used.

Applications for the platform should be written using the supplied SDKs. Further data analysis and business process management also go through the tools provided by the platform itself. The platform has an extensive developer section with instructions, tutorials, and assistance from specialists from the company itself to install, configure, and expand the platform. Also "out of the box" is the possibility of connecting to Microsoft Azure IoT Hub.

\subsection{Public fog platforms} \label{Public fog platforms}
\tab{tab-public-fog}{Overview of public fog platforms}{
\centering
\resizebox{\textwidth}{!}{\begin{tabular}{|l|c|c|c|c|c|} 
\hline
\multicolumn{1}{|c|}{\textbf{ Feature  }} & \textbf{ \makecell{AWS \\ Greengrass}  } & \textbf{ Azure IoT  } & \textbf{ Google  } & \textbf{ Yandex  } & \textbf{ Mail.ru  } \\ 
\hline
\textbf{ \makecell[l]{Supported \\hardware platforms}  } & Universal & Universal & Universal & Universal & Universal \\ 
\hline
\textbf{ \makecell[l]{Basic \\development technology}  } & Universal 
  (Docker) & Universal 
  (Docker) & Universal & Universal & Universal \\ 
\hline
\textbf{ \makecell[l]{Open communication\\ protocols and SDK}  } & + & + & + & + & + \\ 
\hline
\textbf{ Deployment technology  } & Docker & Docker & Docker & Docker & Docker \\ 
\hline
\textbf{ Integration capability  } & \makecell{Amazon Elastic\\Compute 2} & \makecell{Azure,\\via an API} & \makecell{Services of Google\\and partners,\\through API} & \makecell{Universally\\via API} & \makecell{Universally\\via API} \\ 
\hline
\textbf{ \makecell[l]{Connecting external \\data sources}  } & – & – & + & + & – \\ 
\hline
\textbf{\makecell[l]{Availability of\\ additional services \\(Machine Learning, \\Analytics, etc.).  }} & + & + & + & + & + \\ 
\hline
\textbf{ Support Edge  } & + & + & + & + & + \\
\hline
\end{tabular}}
}
Today, public cloud platforms are the solutions of major players in the cloud computing market, focused on solving data processing tasks from IoT systems linked to the capabilities of the corresponding cloud platform. The key characteristics of the considered public fog platform are given in \tabref{tab-public-fog}. 

\textbf{The Azure IoT platform} provides a platform for fog and edge computing based on Microsoft's technology stack. It consists of several extensive subsystems such as IoT Central and IoT Edge, which base their work on Microsoft Azure cloud technology. Connection of devices from Microsoft partners is possible without using drivers or software code due to IoT Plug and Play technology. This approach is possible for devices running any OS, including Linux, Android, Azure Sphere OS, Windows IoT, RTOS, and others. 

Creation, installation, and management of fog applications are performed through the Azure IoT Hub portal. The IoT Hub is a cloud-based managed service that acts as a central message processor for bidirectional communication between an IoT application and the devices it manages. IoT Hub supports both device-to-cloud and cloud-to-device transfers. The IoT Hub supports multiple messaging templates, such as telemetry between devices and the cloud, downloading files from devices, and query-answer technology for managing devices from the cloud. 

To deploy computing closer to the devices themselves or on the devices themselves, uses Azure IoT Edge, which allows you to deploy applications with their business logic, or already available in the directory ready-made applications on end devices using containerization technology. 

\textbf{The Amazon AWS IoT Greengrass platform} allows you to extend the capabilities of AWS (Amazon Web Services) to your peripherals, enabling them to work locally with your data while using the cloud to manage, analyze and securely store your data. AWS IoT Greengrass allows connected devices to perform AWS Lambda functions, run Docker containers, generate forecasts based on machine learning models, synchronize these devices and interact securely with other devices even without an Internet connection. 

AWS IoT Greengrass allows you to create IoT solutions that connect different types of devices to the cloud and each other. AWS IoT Greengrass Core can be used on Linux devices (including Ubuntu and Raspbian distributions) that support Arm or x86 architectures. The AWS IoT Greengrass Core service provides local AWS Lambda code execution, messaging, data management, and security. Devices with AWS IoT Greengrass Core serve as portals of the service and interact with other devices that run FreeRTOS (Real-time operating system for microcontrollers) or installed SDK package AWS IoT for devices. The size of such devices can be very different: from small devices based on microcontrollers to large household appliances. When a device with AWS IoT Greengrass Core loses contact with the cloud, devices in the AWS IoT Greengrass group can continue to communicate with each other over a local network. 

Google, Yandex, and Mail.ru platforms provide their cloud and fog solutions for data collection, storage, processing, analysis, and visualization. Collected data from devices is integrated into the public cloud system for deeper processing and analysis (including machine learning and artificial intelligence) due to the high computing power of the cloud. These platforms support multiple protocols for connectivity and communication through the provided API. There are many ready-to-use services available for installation in the platform directory itself, which can be connected to your cloud solution by combining them.

\subsection{Open Source Fog Platforms} \label{Open Source Fog Platforms}
During the analysis of existing solutions, we reviewed existing open-source fog platforms. In contrast to commercial solutions, for open-source platforms, there are complete descriptions of architectures, requirements to computing resources, as well as technologies used, both on hardware and software levels (see \tabref{tab-opensource-fog}).

\textbf{FogFrame2.0 }is an open-source fog platform \cite{FogFrame} aimed at deployment on single-board computers (Raspberry Pi). Authors designed architecture and implemented a representative framework to resolve the following challenges \cite{Skarlat2019}:
\begin{itemize}
    \item enable the coordinated cooperation among computational, storage, and networking resources in the fog \cite{Skarlat2017, Skarlat2016};
    \item implement heuristic algorithms for service placement in the fog, namely a first-fit algorithm and a genetic algorithm;
    \item introduce mechanisms for adapting to dynamic changes in the fog landscape and for recovering from overloads and failures.
\end{itemize}
To evaluate the behavior of FogFrame, authors apply different arrival patterns of application requests, i.e., constant, pyramid, and random walk, and observe service placement. The platform dynamically reacts to events at runtime, i.e. when new devices appear or are disabled when devices experience failures or overloads, necessary node redeployments are performed.  

\textbf{The FogFlow platform} is an open-source fog platform \cite{FogFlow}. The developers' main task was to provide a flexible and straightforward way of development, deployment, and orchestration of fog services \cite{Bonomi2012}. The uniqueness of their approach is in the following: 
\begin{itemize}
    \item standard-based programming model for fog computing with declarative hints; 
    \item scalable context management: to overcome the limitations of centralized context management, FogFlow introduces a distributed context management approach.
\end{itemize}
The data structure of all data flows is described based on the same standardized data model called NGSI. Therefore, FogFlow can learn which type of data is created at which edge node. It then triggers and launches dynamic data processing flows for each edge node based on the availability of registered context metadata, which gives service developers two advantages:
\begin{itemize}
    \item fast and easy development of fog computing applications, because the proposed hints hide configuration and deployment complexity tasks from service developers;  
    \item good openness and interoperability for information sharing and data source integration with the use of NGSI - a standardized open data model and API and it has been widely adopted by more than 30 cities worldwide.
\end{itemize}
FogFlow is one of the components of FIWARE open infrastructure \cite{FIWARE2015}, which provides the development and implementation of various smart solutions \cite{Celesti2019, D.2011, Fazio2016}. This infrastructure is one of the modern cloud frameworks along with Amazon Web Services \cite{Guth2017}. A wide library of ready-made solutions from the developer community and detailed implementation instructions are available for implementation and use of FogFlow \cite{FogFlowDocumentation}.

\textbf{The FogBus platform} (supported by Melbourn Clouds Lab) integrates various hardware tools through software components that provide structured interaction and platform-independent application execution \cite{Tuli2019}. FogBus uses blockchain to ensure data integrity when transmitting sensitive data. The platform-independent architecture of application execution and interaction between nodes allows overcoming heterogeneity in the integrated environment.  

FogBus supports implementing various resource management and scheduling policies to run IoT applications compiled using parallel programming models such as SPMD (single program, multiple data). 

To evaluate the performance of the FogBus platform, a prototype application system is used to analyze the Sleep Apnea data. This example illustrates how an application (in the healthcare sector) built using the SPMD model can be implemented using different FogBus settings to process IoT data in an integrated computing environment. 

This framework makes it easy to deploy IoT applications, monitor and manage resources. FogBus system services are developed in cross-platform programming languages (PHP and Java). Thay are used with the Extensible Application Layer Protocol (HTTP), which helps FogBus overcome heterogeneity in the communication level of the OS and P2P of different nodes of fog. Besides, the FogBus platform functions as a "Platform as a Service" (PaaS) model for the Fog Cloud integrated environment, which not only helps application developers create different types of IoT applications but also supports users to configure services, and service providers to manage resources according to system conditions.

\tab{tab-opensource-fog}{Overview of Open Source Fog Platforms}{
\centering
\resizebox{\textwidth}{!}{\begin{tabular}{|l|l|c|} 
\hline
~ & \multicolumn{1}{c|}{\textbf{ Goal  }} & \textbf{ Deployment  } \\ 
\hline
\textbf{ FogFrame2.0  } & Check
  the conceptual model & – \\ 
\hline
\textbf{ FogFlow  } & Simpler
  and more flexible orchestration of services & + \\ 
\hline
\textbf{ FogBus  } & \begin{tabular}[c]{@{}l@{}}Overcome heterogeneity at the communication \\level between OS and P2P of different nodes of the fog\end{tabular} & – \\
\hline
\end{tabular}}
}

\subsection{Classification methods for fog platforms} \label{Classification methods for fog platforms}
To form a unified approach to the classification of fog platforms, we considered the key fog platforms and their key characteristics. For example, AWS Greengrass can work without access to the public cloud\footnote{\url{https://aws.amazon.com/ru/greengrass/faqs/Local_Resource_Access}}, but it is possible only to store local data in this mode of operation. Central device management, as well as centralized data collection and processing, becomes impossible. Entire platform operation requires access to AWS IoT Core, which acts as a central service for the management and organization of fog and a public cloud.

Azure IoT can also operate on private networks\footnote{\url{https://azure.microsoft.com/en-us/blog/introducing-iot-hub-device-streams-in-public-preview/}}, but only if there is a gateway within the private network that must connect to the central management and data collection node, and that node is also a public cloud. What distinguishes IoT from a public cloud is that it has a single point of access to the external network, rather than many different gateways that communicate with the public cloud. 

Other public fog platforms have the same limitations as private and open-source fog platforms, the central control node of which can be deployed on any server in the local network or not at all (in this case, management and orchestration tasks are separated between intermediate fog nodes, as is done, for example, in FogFlow2.0). 

Therefore, all fog platforms can be classified according to the openness or closedness of the deployment of the hub, a service that is responsible for connecting, monitoring and managing devices connected in the fog. In one form or another, almost all commercial fog platforms has the hub: LoopEdge and Azure IoT call this service - the Hub. ClearBlade and FogHorn platforms have a service with the same functionality, but it is called Device Manager. At AWS Greengrass, this service is called AWS IoT Core. 

Another criterion for classification may be the requirements for the underlying hardware on which fog platform services can be deployed. Some of these platforms are tightly bound to a limited list of supported hardware devices. On the other hand, other platforms allow their services to be deployed on any end-user hardware as long as it meets the necessary minimum requirements for platform deployment. We define this characteristic as an indicator of the classification of fog platforms according to the openness of the hardware infrastructure. 

The same principle is observed when comparing platforms based on openness or closed software infrastructure: the platform can support open protocols of data exchange between nodes of fog or fog programs are supplied exclusively by the developers of the platform itself and licensed partners. 

Thus, any fog platform can be classified according to the principle of openness or closeness of its components (see Fig. 2). It should also be noted that platforms with a public hub are more likely to be open to their hardware and software infrastructures.

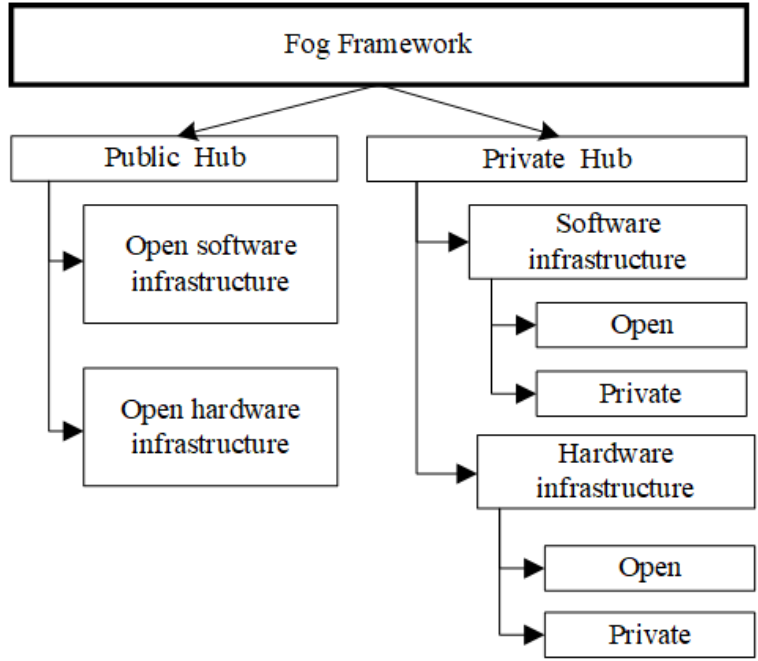
\fig{width=0.6\textwidth}{pic/Fig-Classification-1}{Classification of fog platforms according to the principle of openness or closeness of its components}

In addition to the openness or closeness of their components, some platforms have focused on the availability of the various features or services provided by the platform. The Azure IoT Hub, which is an integral part of the Azure IoT platform, explicitly calls itself \textbf{PaaS (Platform as a Service)}, providing ready-made solutions for the user's required tasks. It should be noted that none of the public fog platforms positions their platforms as pure fog. They provide fog computing as a certain basic functionality, which is the basis for other provided platform functions and services. 

Thus, the platforms themselves position some functionality as basic, which should be in any fog platform, and the user is interested not only in simple deployment and basic management of fog nodes but also in solving their specific tasks: Industry 4.0, Medicine, Smart City, etc. Platforms should provide ready-made solutions for each of the user's tasks as much as possible. 

Among other things, some platforms have allowed users to share their ready-made solutions created within the platform with the help of "stores" - resources where the user can publish his readymade fog application. This has led to the emergence of entire fog ecosystems -\textbf{ EaaS (Ecosystem as a Service)}, which allow users to create their fog solution from ready-made components available on the platform. 

This description also includes Open Source solutions that provide only a basic level of functionality - \textbf{DaaS (Deploy as a Service)}: deployment of fog nodes on existing devices, orchestration, etc. On the other hand, FogFlow has wider functionality and even its ecosystem, which includes ready-to-install components from both platform developers and the community. 

\textbf{Classification "as a service"} can be used as a classification method based on the provided platform functionality (see \figref{pic/Fig-Classification-2}).

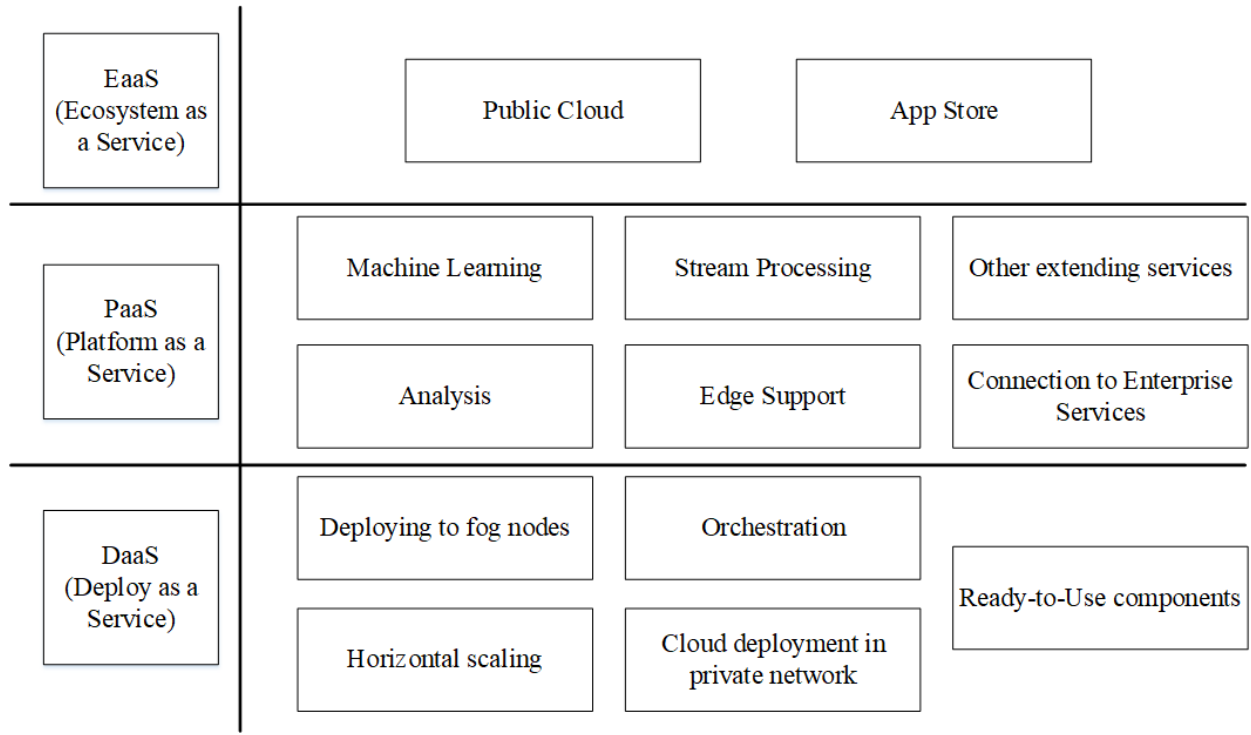
\fig{width=1\textwidth}{pic/Fig-Classification-2}{Classification of fog platforms based on provided platform functionality}

\section{Fog computing challenges} \label{Challenges}

In this section, we discuss several future research directions that are considered to be most promising for future research in other works and this research likewise. 

\textbf{Artificial Intelligence Application Management} is currently receiving considerable attention because of its ability to solve complex problems. The data needed to build an AI system is quickly accumulated in fog \cite{Li2019}. Artificial Intelligence application management can help predict future resource requirements, context variations, and node failures more accurately and manage applications accordingly.

Fog nodes are limited in resources. Adding more fog nodes to the fog may reduce this limitation. However, it increases the cost of deployment, complexness of node communication, and power consumption at the network edge \cite{Afrin2019}. In this case, it may be helpful to \textbf{dynamically consolidate and scale the fog nodes} according to computational needs.
Fog computing is developed to execute various complex IoT applications from different domains, including smart healthcare, city, agriculture, and industry \cite{Mohamed2019}. These IoT applications have specific requirements and need specialized support. \textbf{Application-specific management strategies} can help deal with them in Fog.

Task and data processing is decentralized in the Fog. The task may begin on one node and going through several others end on the last one. When an emergency happens in the fog, the developer and fog designer need access to log information to locate the problem in minimal time. Thus total logging helps with this task, but then the problem appears to maintain a data lake supporting storage and analysis of such data. That’s the question of \textbf{logging and monitoring of highly-distributed fog applications}.

On the other hand, there is also the issue of \textbf{task sharing and re-usability}. Applications can share a particular task to optimize the computational load on fog nodes \cite{Varshney2017}. Besides, the task executables of recently terminated applications can also be reused for other applications. To perform such operations, shared caching techniques and policies are required to be developed in the context of fog computing. 

The above opens another question. If the fog node faces a software or hardware problem and shuts down other nodes won’t have any information on the checkpoint the node was in. But most applications are state-dependent and stateful. So there is a challenge to organize \textbf{state management and sharing} between nodes to support the continuous and flowless work of the fog.

Most Fog applications do not consider security as part of a system but rather focus on functionality, which results in many fog platforms being vulnerable \cite{Khan2017}. That leads to sensitive data leakage, user loss of privacy, and other \textbf{security issues} that are very significant in most IoT domains. Future work could lead towards the development of knowledge-based supplementary references, which can provide decision support for developers in designing a secure and performance efficient fog infrastructure. Such decision support would require a large systematic knowledge acquisition of best practices, known security threats and their solutions, which can be formalized as either a statistical-based system or rules, policies and facts.

\section*{Conclusion} \label{Conclusion}
The increase of transferred data volumes and the increased load on the cloud for client services became a prerequisite for the concept of cloud computing. In this paper, the concept of fog computing, its definition and key characteristics were considered. Also, there were considered, classified and generalized some fog platforms, which are subjects of research or already used by business and private clients. In the end, the general architectural characteristics inherent in all the platforms reviewed were described. 
Fog computing is a more flexible and efficient type of computation compared to cloud computing due to the solution of tasks requiring high bandwidth of the computing network, the ability to work with geographically dispersed data sources, ultra-low latency and providing local data processing. 

In this review paper, we not only to given an extended point of view over the fog computing paradigm but to also analyzed the growing diverse number of open source and enterprise solutions for deploying fog platforms. On the basis of this review, we proposed a classification of fog solutions by their cloud layer, hardware and software publicity level and by a provided service and functionality they grant.

\section*{Acknowledgements}
The research was carried out with the financial support of RFBR and Chelyabinsk Region, project number 20-47-740005 and the Ministry of Science and Higher Education of the Russian Federation (state assignment FENU-2020-0022).

\openaccess

\bibliography{References}
\end{document}